\documentclass[twocolumn]{aastex6}
\newcommand{\asat}{{\em AstroSat}}
\newcommand{\xray}{X--ray}
\newcommand{\element}[2][]{\ensuremath{^{#1}\texttt{#2}}}
\newcommand{\us}{\ensuremath{\mu{\rm s}}}
\newcommand{\degdot}{\degr\hspace{-0.3em}.}

\slugcomment{To appear in JAA special issue on Astrosat}
\shorttitle{Cadmium Zinc Telluride Imager}
\shortauthors{Bhalerao et al}

\begin{document}

\title{The Cadmium Zinc Telluride Imager on Astrosat}

\author{V. Bhalerao\altaffilmark{1,2}, D. Bhattacharya\altaffilmark{2}, A. Vibhute\altaffilmark{2}, P. Pawar\altaffilmark{2,3}, 
A. R. Rao\altaffilmark{4}, M. K. Hingar\altaffilmark{4}, Rakesh Khanna\altaffilmark{4}, A. P. K. Kutty\altaffilmark{4}, J. P. Malkar\altaffilmark{4},
M. H. Patil\altaffilmark{4},   Y. K. Arora\altaffilmark{4}, S. Sinha\altaffilmark{4}, 
P. Priya\altaffilmark{5},  Essy Samuel\altaffilmark{5}, S. Sreekumar\altaffilmark{5}, P. Vinod\altaffilmark{5},
N. P. S. Mithun\altaffilmark{6}, S. V. Vadawale\altaffilmark{6}, N. Vagshette\altaffilmark{6,2}, 
K. H. Navalgund\altaffilmark{7}, K. S. Sarma\altaffilmark{7}, R. Pandiyan\altaffilmark{7}, S. Seetha\altaffilmark{7},  K. Subbarao\altaffilmark{7}}

\altaffiltext{1}{Email: \url{varunb@iucaa.in}}
\altaffiltext{2}{Inter University Centre for Astronomy \& Astrophysics, Pune, India}
\altaffiltext{3}{S. R. T. M. University, Nanded, India}
\altaffiltext{4}{Tata Institute of Fundamental Research, Homi Bhabha Road, Mumbai, India}
\altaffiltext{5}{Vikram Sarabhai Space Centre, Thiruvananthapuram, India}
\altaffiltext{6}{Physical Research Laboratory, Ahmedabad, India }
\altaffiltext{7}{ISRO Satellite Centre, Bengaluru, India}

\begin{abstract}
The Cadmium Zinc Telluride Imager (CZTI) is a high energy, wide--field imaging instrument on \asat. CZTI's namesake Cadmium Zinc Telluride detectors cover an energy range from 20~keV to $>200$~keV, with 11\% energy resolution at 60~keV. The coded aperture mask attains an angular resolution of 17\arcmin\ over a 4\degdot6 $\times$ 4\degdot6 (FWHM) field of view. CZTI functions as an open detector above 100~keV, continuously sensitive to GRBs and other transients in about 30\% of the sky. The pixellated detectors are sensitive to polarisation above $\sim100$~keV, with exciting possibilities for polarisation studies of transients and bright persistent sources. In this paper, we provide details of the complete CZTI instrument, detectors, coded aperture mask, mechanical and electronic configuration, as well as data and products.
\end{abstract}


\section{Introduction} \label{sec:intro}

\begin{figure}[!tbh]
\includegraphics[width=\columnwidth]{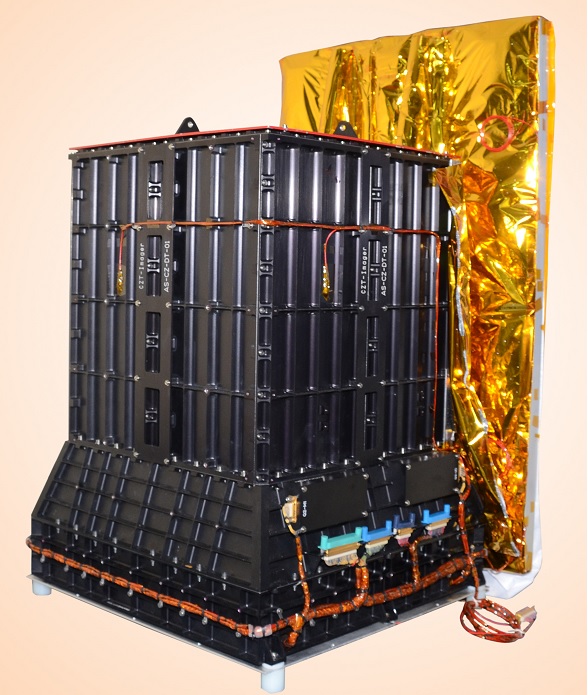}
\caption{The fully assembled Cadmium Zinc Telluride Imager before integrating with the spacecraft.}
\label{fig:cztisideview}
\end{figure}

\asat, India's first space observatory, was launched into low earth orbit on 28 September 2015. \asat\ carries four co--pointed instruments giving broadband coverage in optical, ultra--violet, soft \xray\ and hard \xray\ bands, as well as a soft \xray\ all--sky monitor~\citep{astrosat}. Here, we describe the Cadmium Zinc Telluride Imager (CZTI\footnote{More details about the Cadmium Zinc Telluride Imager can be found at \url{http://astrosat.iucaa.in/czti}.}), the wide--field instrument providing the hard \xray\ coverage for \asat~(Figure~\ref{fig:cztisideview}).

CZTI was conceived with the aim of using modern solid state detectors to bring good sensitivity in hard \xray s to \asat. The guiding design principles were to have approximately thousand square centimetres collecting area, high sensitivity extending above 100~keV, imaging capabilities, and a wide field of view to obtain good background measurements. The final design driven by these considerations consists of an array of CZT detector modules sensitive to 20--200~keV \xray s, coupled with a coded aperture mask to obtain 8\arcmin\ resolution for bright sources over a 4\degdot6 $\times$ 4\degdot6 field of view.

\begin{figure*}[htbp]
\plottwo{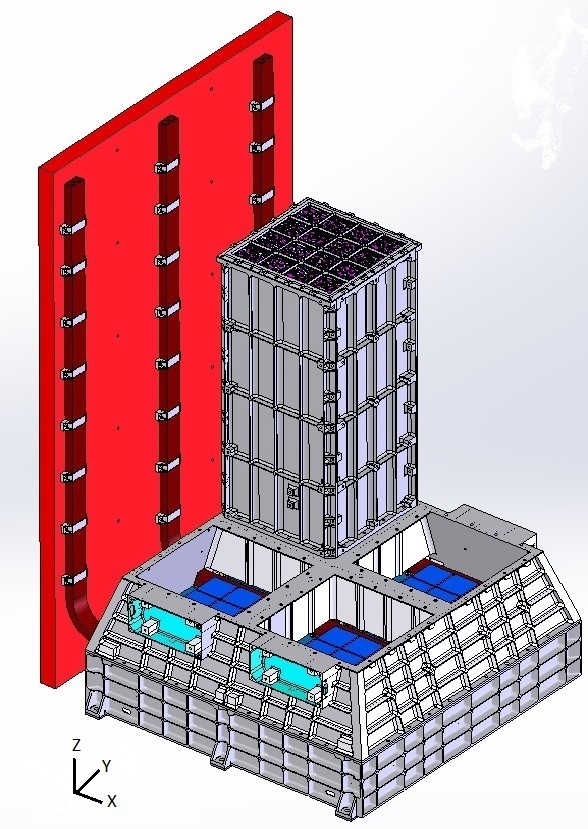}{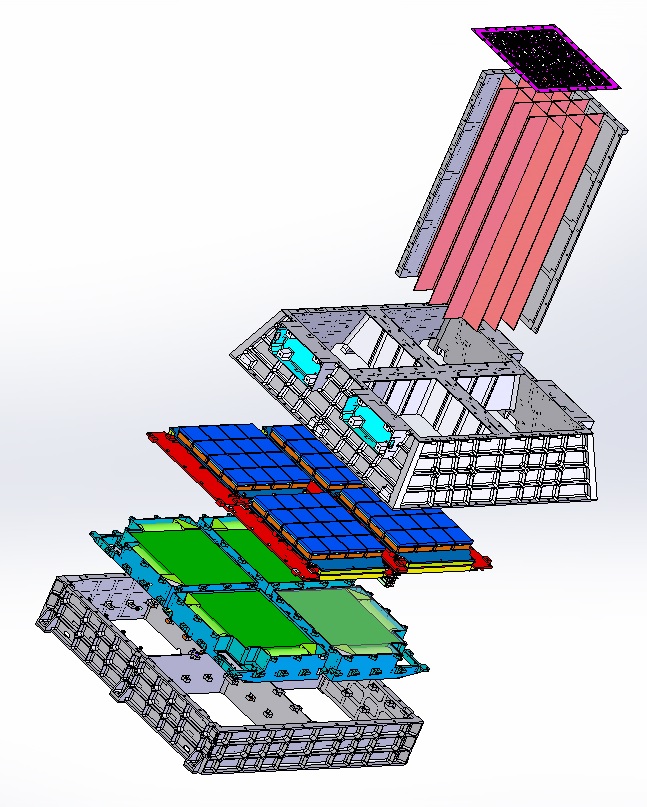}
\caption{Layout of the Cadmium Zinc Telluride Imager. 
\textit{Left Panel:} The instrument consists of four identical quadrants with coded aperture masks at the top. The collimator housing for three quadrants is hidden for clarity. CZT detectors (blue) are located in the detector housing. The spacecraft co--ordinate system is marked in the lower left, with $Z$ axis oriented with the instrument bore sight. A radiator plate (red) is located along the $-X$ axis, along with side coding masks on the detector housing. Alpha--tagged \element[241]{Am} sources (cyan) are placed near the CZT detectors, in $\pm Y$ directions.
\textit{Right Panel:} An exploded view of CZTI to show all components. The Coded Aperture Mask is located at the top. Two walls of the collimator housing (grey) are hidden to show the collimators (salmon / pink) aligned with each module. Sixteen CZT modules (blue) in each quadrant are mounted on an electronics board (red). \element{CsI}(\element{Tl}) veto detectors (green) located below the CZT modules serve as active anti--coincidence shields to reject charged particle events and very high energy photons. \element[241]{Am} alpha--tagged radioactive sources (cyan) are mounted alongside the detector housing are used to monitor the CZT module energy response in orbit. }
\label{fig:cztilayout}
\end{figure*}

This hard \xray\ coverage, coupled with the broadband abilities of other instruments, allows \asat\ to address a wide range of scientific problems. Science objectives of CZTI include the measurement of curvature and reflection components in the spectra of Active Galactic Nuclei and \xray\ binary systems, the study of Quasi-Periodic Oscillations at hard \xray\ bands in accreting neutron star and black hole systems, cyclotron line spectroscopy of high mass \xray\ binaries, the characterisation of hard \xray\ spectra of magnetars as well as the detection of gamma ray bursts and the study of their early light curves. 

\xray\ polarimetry is an important yet unexplored area in high energy astrophysics. Polarisation carries important information about the emission mechanisms at the source, as well as physical conditions and geometry responsible for the emission. If polarised \xray\ photons undergo Compton scattering in a detector, they are preferentially scattered perpendicular to the direction of polarisation. This photon may then interact in a neighbouring pixel giving a near--simultaneous signal with the primary interaction. A significant fraction of 100~keV--300~keV \xray s undergo Compton scattering in CZTI, creating two--pixel events. We can exploit the directionality of double events to measure polarisation of bright \xray\ sources and GRBs. With this unique capability, CZTI can securely measure $>40\%$ polarisation at 3-$\sigma$ level for bright sources (500~mCrab) in just 100~ks exposures \citep{cvr+14,vcr+15}.

The layout of this paper is as follows. In \S\ref{sec:mech} we discuss the overall instrument configuration and the coded aperture mask. The detectors and electronics are discussed in \S\ref{sec:detelec}. Instrument data, processing pipelines and default products are discussed in \S\ref{sec:data}. We conclude by giving brief information about in--orbit performance of the instrument in \S\ref{sec:inorbit}.

\section{Instrument configuration}\label{sec:mech}

The instrument consists of four identical independent quadrants A--D, to give design safety and redundancy (Figure~\ref{fig:cztilayout}). Each quadrant consists of a Coded Aperture Mask (CAM; \S\ref{subsec:cam}) at the top. Sixteen CZT detector modules (\S\ref{subsec:modules}) are mounted at a distance of 48~cm below the CAM. Each quadrant also has an alpha--tagged \element[241]{Am} source for gain calibration just above and to the side of the detector plane (\S\ref{subsec:alpha}), and a \element{CsI} (\element{Tl}) veto detector below the CZT modules (\S\ref{subsec:veto}). Each quadrant has an independent Front end Electronics Board (FEB; \S\ref{subsec:electronics}) for providing power and processing all signals. All the four quadrants are connected to common Processing Electronics which act as the interface to the satellite. The overall dimensions of the payload are 482~mm $\times$ 458~mm $\times$ 603.5~mm. In addition to the four quadrants, the instrument also has a radiator plate affixed on a side to dissipate heat and provide a cold bias for maintaining CZT module temperatures.

Next, we discuss the structural elements critical for source localisation with CZTI.

\subsection{Coded Aperture Mask}\label{subsec:cam}
\begin{figure}[htbp]
\includegraphics[width=\columnwidth]{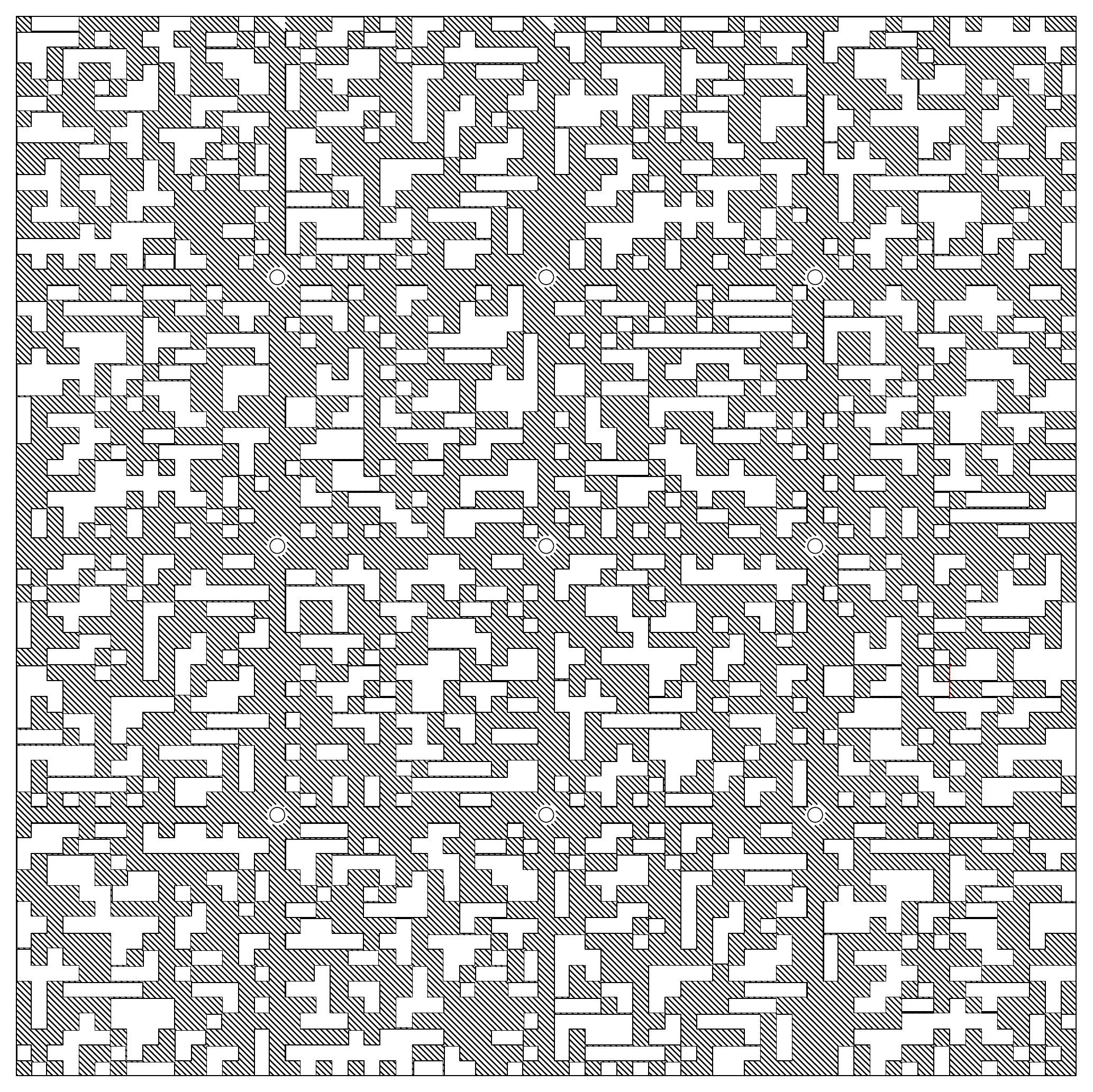}
\caption{Top view of the coded aperture mask for Quadrant~A.
The horizontal and vertical axes in this figure are aligned to the instrument $X$ and $Y$ axes respectively. The holes are about 2.46~mm on a side, and the edge holes for each module are 2.31~mm to match the pixel sizes of the CZT detector modules. Neighbouring masks are separated by 2.50~mm strips to mimic the inter--module separation in the detector plane. The mask patterns for quadrants B, C, D are obtained by rotating this pattern clockwise by 90\degr, 180\degr\ and 270\degr\ respectively. Dimensions marked are in mm.}
\label{fig:cam}
\end{figure}

A Coded Aperture Mask (CAM) is basically a mask with a pattern of open and closed squares/rectangles that cast a unique shadow on the detector plane for each source direction~\citep{csc+87}. 
The CZTI CAM is made of a 0.5~mm thick Tantalum plate with square and rectangular holes matching the pitch of the CZT detector pixels. Additional support bridges of thickness 0.2~mm are introduced at a number of places within the pattern to improve its mechanical stability. 
The patterns are based on 255-element pseudo-noise Hadamard Set Uniformly Redundant Arrays. Of sixteen possible such patterns, seven were chosen on the basis of the mechanical support for individual pixels in the pattern. These seven patterns, with some repeats, were placed in the form of a 4 x 4 matrix to generate the CAM for Quadrant~A (Figure~\ref{fig:cam}). This same pattern is placed on other quadrants, rotated clockwise by 90\degr, 180\degr\ and 270\degr\ respectively. 

The CAM is mounted horizontally on top of a collimator housing made from Aluminium alloy (AA6061T6). 
The detector plane is 481~mm below the lower surface of the CAM. 
The inter--detector boundaries are lined with vertical collimator plates extending for a distance of 400~mm below the CAM. The Collimator plates are made of 0.07 mm thick Tantalum sheet mounted on 2~mm thick precisely machined Aluminium alloy plates, and terminate 8~cm above the detector plane to accommodate the calibration housing. These plates prevent illumination of any detector through the mask of a nearby detector. 

The simple box--type design of the CZTI CAM\footnote{A `box--type' system is such that the size of the mask plate is the same as that of the detector itself.} and the collimators together provide a collimation of $4\degdot6 \times 4\degdot6$ FWHM. However due to the 78~mm gap between the bottom of the collimator and the detector plane, a certain amount of illumination leakage occurs from one collimated module to its adjacent neighbours. When this illumination overlap is taken into account, the net field of view works out to be $11\degdot8 \times 11\degdot8$ (Full Width at Zero Maximum). The typical 2.46~mm mask elements yield a geometric angular resolution of 17\arcmin. 

\subsection{Side coding}\label{subsec:sidecoding}
\begin{figure}[thbp]
\centering
\includegraphics[width=0.9\columnwidth]{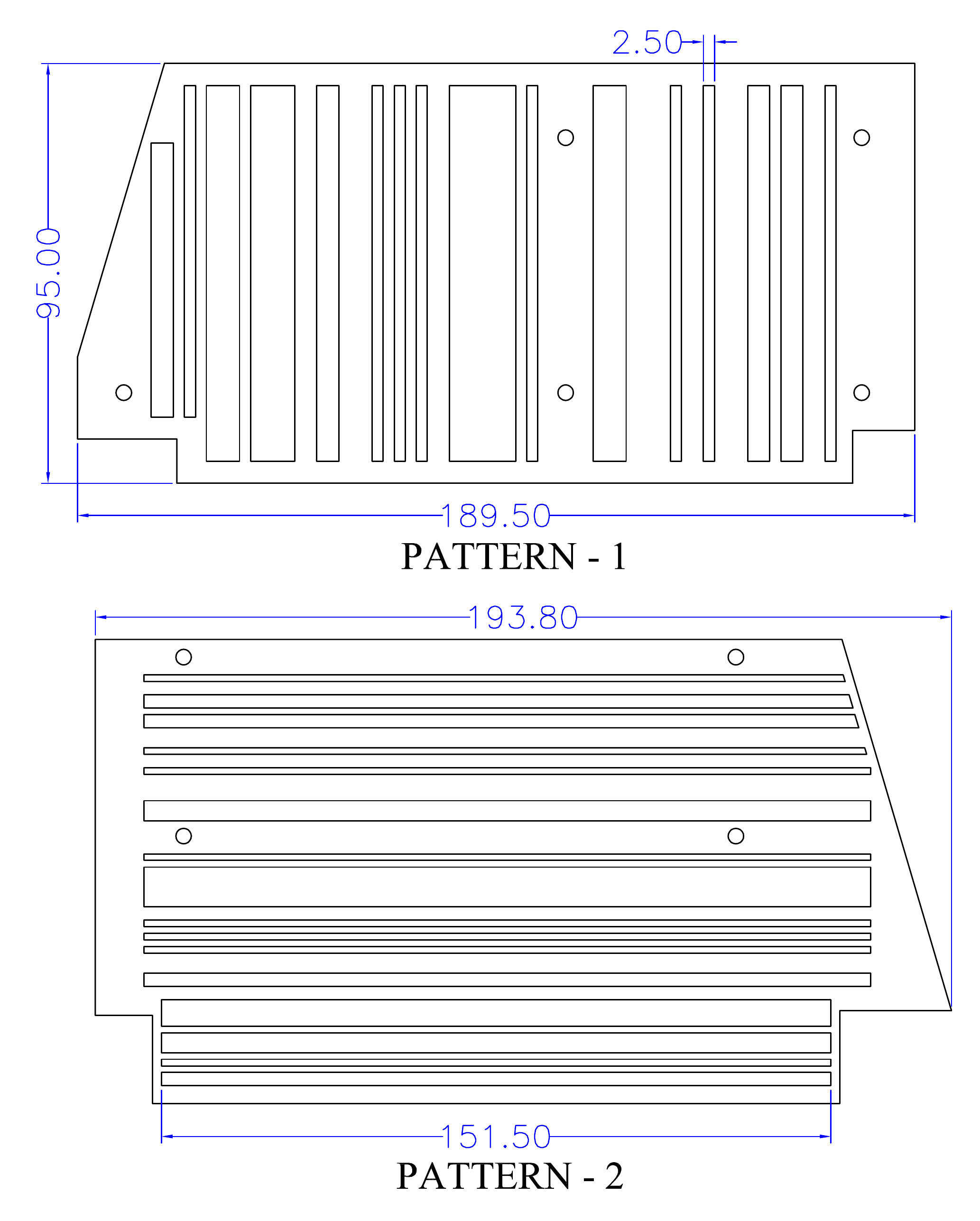}
\caption{The side coding mask in CZTI. The mask is located between the radiator plate and the CZT modules, inclined at 18\degr\ to the vertical. Top: the mask located next to Quadrant~A has vertical slits with a 2.5~mm pitch. Bottom: the mask located next to Quadrant~D has horizontal slits with a 1.5~mm pitch.}
\label{fig:sidecam}
\end{figure}

To aid localisation of sources located off axis, CZTI carries a set of one dimensional coded patterns on one side of the calibration housing. This mask is attached to the underside of the walls of the housing between the radiator plate and the detectors, inclined at an angle of 18\degr\ from the vertical (Figure~\ref{fig:cztilayout}). Thus, normal incidence on these patterns occurs for sources located 72\degr\ away from the instrument bore--sight in the $-X$ direction. 
The mask is made of 0.5 mm thick Tantalum plate, which is bonded to the 1.5 mm thick Aluminum housing wall. An additional shielding of 0.07~mm thick Tantalum is also provided on this surface. 
The radiator plate and the housing wall block lower energy photons, and the side coding is most effective in the 60--200~keV range. The mask next to Quadrant~A has vertical slits with 2.5~mm pitch, while the part next to Quadrant~D has 1.5~mm pitch horizontal slits (Figure~\ref{fig:sidecam}). Both mask patterns are designed from 63--element pseudo-noise Hadamard Uniformly Redundant Arrays, as in \asat\ Scanning Sky Monitor~\citep{astrosat}.

The side coding effective area peaks around 300~keV, with the maximum effective area projected in the direction of the source being about 140~cm$^2$ within the side coded field of view. The positional accuracy provided by the side coding is about a degree for a source with about a thousand counts.

\subsection{Off axis response}\label{subsec:offaxis}

\begin{figure}[!thbp]
\includegraphics[width=\columnwidth]{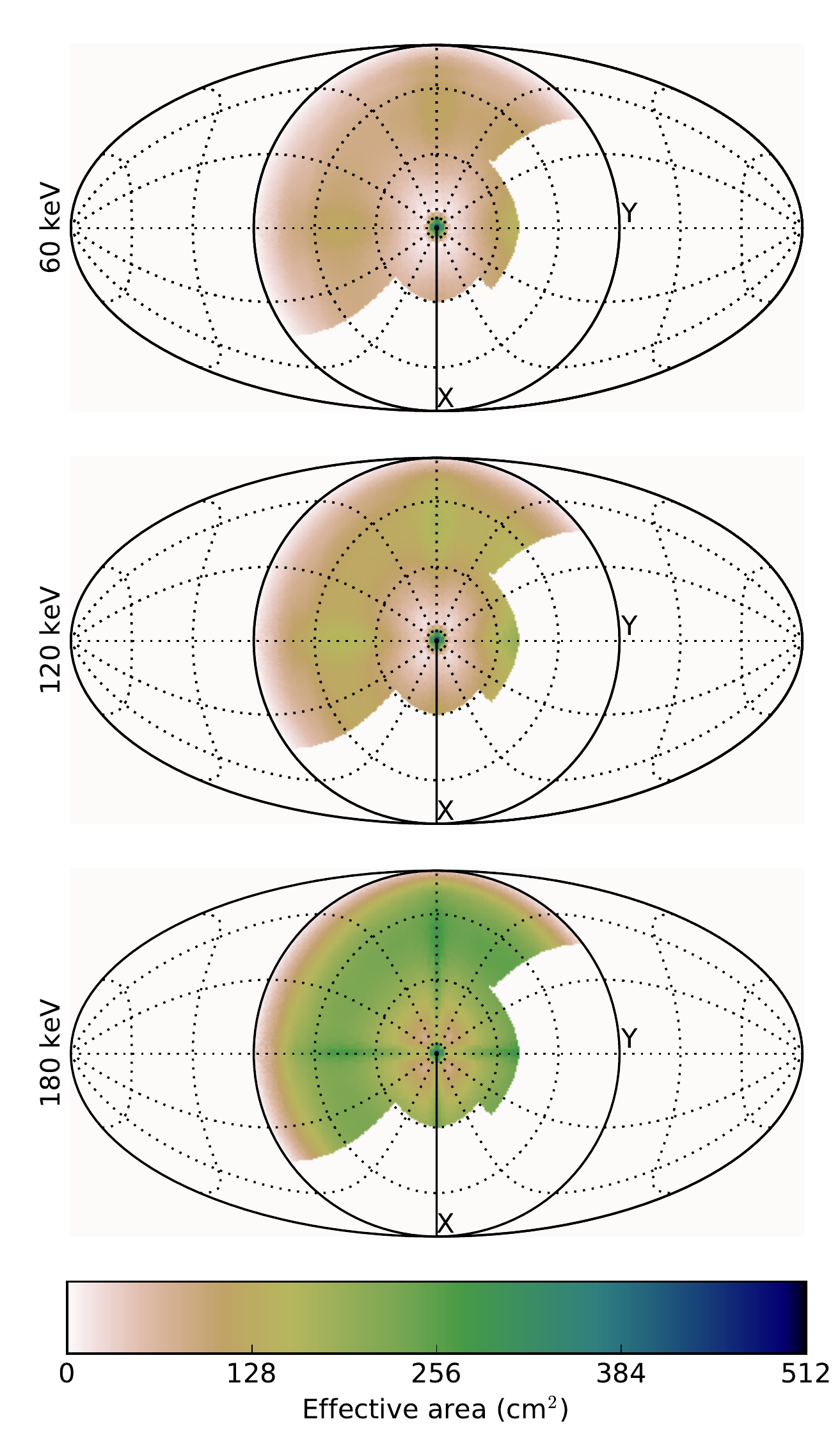}
\caption{Off-axis effective area of CZTI at various energies. The maps are in a mollweide projection, with the CZTI pointing direction ($Z$ axis) at the centre of the map. The $X$ and $Y$ axes are as in Figure~\ref{fig:cztilayout}. For this figure, effective area has been set to zero for lines of sight below the detector plane, or obstructed by LAXPC ($+X$ region) or SXT ($+Y$ region). The median effective area at 180~keV is 190~cm$^2$.}
\label{fig:offaxis}
\end{figure}

The housing of CZTI, being made of Aluminium alloy and thin Tantalum shields, is transparent above $\sim100$~keV. This makes CZTI an excellent wide--angle monitor, covering roughly one third of the sky at all times (Figure~\ref{fig:offaxis}). This monitoring capability has been leveraged to detect several transients\footnote{All transients discovered by CZTI are listed at \url{http://astrosat.iucaa.in/czti/?q=grb}.}, including the detection of GRB~151006A on the first day of operation, 60\degdot7 away from the pointing direction \citep{bbr+15}. 
The net transmission of the CZTI housing and collimators depends on the incidence angle and the energy. We utilise this angular dependence for localising transient sources with an uncertainty of just a few square degrees. For instance, \citet{rch+16} apply this technique to localise GRB~151006A accurately to $\sim10\degr$. Further work to improve off--axis localisation capabilities is under way and will be reported elsewhere.

\section{Detector characteristics}\label{sec:detelec}

\subsection{CZT detector modules}\label{subsec:modules}
The principal detectors in CZTI are sixty four Cadmium Zinc Telluride detectors, called detector modules. The modules, purchased from Orbotech Medical Solutions, Israel\footnote{Orbotech Medical Solutions (\url{http://www.orbotech.com/}) is now a subsidiary of GE Medical.}, are made of 5~mm thick crystals of the compound semiconductor \element{Cd}$_{0.9}$\element{Zn}$_{0.1}$\element{Te}$_{1.0}$. Each module has a continuous anode made of 50~$\mu$m aluminised mylar. The cathode is divided into a 16$\times$16 pixel grid, which is directly bonded to two Application Specific Integrated Circuits (ASICs) for readout. The ASIC signals are further read and processed by other electronic systems in the satellite (\S\ref{subsec:electronics}). These modules function with high quantum efficiency in the 20--150~keV range, and with lower efficiency to energies above 200~keV. The detection threshold (also known as LLD, Lower Level Discriminator) can be set individually for each of the 64 modules in the instrument. With our current LLD and gain settings, most modules in CZTI are sensitive to an energy range approximately from 15 to 200 keV. Further details about the modules are given in Table~\ref{tab:cztdet}.

Ground calibration tests for all pixels in all modules show that the energy response is linear to better than one percent. In--orbit measurements of the energy--PI gain and energy resolution can be carried out using an onboard radioactive source (\S\ref{subsec:alpha}). The energy resolution is 11\% at 60~keV, and has not changed between ground calibrations and in--flight. Further details of the module response and energy profiles are discussed in \citet{cvr+16}.

The low noise ASICs enable operation of these modules at room temperature, significantly simplifying the testing and calibration of the instrument. In space, the module temperature is maintained by using heaters and cold fingers connected to a radiator plate. The radiator plate always points in the anti--sun direction, and gets passively cooled to --30\degr C. Heating elements are placed on the Internal Radiator Plate to maintain the module temperature in the 10\degr C -- 15\degr C range.

In addition to imaging and spectroscopy, the CZT modules also have good timing abilities. The response time of individual CZTI pixels is $<1~\us$, with negligible dead time. There are no pileup effects even for bright sources\footnote{Pileup can occur if count rates become comparable to $\sim10^6$ counts/pixel/sec. For a nominal 2.46~mm pixel, this corresponds to $\gtrsim 10^7{\rm~counts~sec~cm}^{-2}$ in the entire CZTI band. For comparison the Crab gives approximately 0.08 ${\rm~counts~sec~cm}^{-2}$ in the 15--200~keV range}. Detected events are stored in an internal First--In--First--Out (FIFO) data buffer, which are read out by the Front end Electronics Board (\S\ref{subsec:electronics}). Based on expected count rates from bright sources, we fixed the instrument time resolution to 20~\us.

\begin{table}[htbp]
\caption{CZT detector properties}
\label{tab:cztdet}
\begin{tabular}{p{0.45\columnwidth}p{0.5\columnwidth}}
\hline
Property & Value \\
\hline
Physical properties & \\
\hline
Composition  & Cd$_{0.9}$Zn$_{0.1}$Te$_{1.0}$\\
Growth method & MHB (Modified Horizontal Bridgman)\\
Conductivity type  & N-Type\\
Density  & 5.85 g~cm$^{-3}$\\
Bulk resistivity  & $3 - 8 \times 10^9~\Omega$~cm \\
Pair creation energy (eV)  & 4.43~eV\\
Electrode type  & Ohmic\\
Electrode material  & Indium\\
$\mu\tau$ (Electron mobility $\times$ lifetime product) & $3-5\times10^{-3}{\rm~cm}^2~{\rm V}^{-1}$\\
Theoretical absorption  & $> 88\%$ (at 122~keV)\\
\hline
Geometry & \\
\hline 
Length & 39.06 $\pm$ 0.05~mm\\
Width & 39.06 $\pm$ 0.05~mm\\
Thickness & 5.0 $\pm$ 0.1~mm\\
Pixels matrix & $16 \times 16$\\
Pixel size & 2.46~mm for central 14$\times$14 pixels, 2.31~mm for side rows and columns\\
Pixel anode pad size & $1.86 \times 1.86$~mm\\
\hline
\end{tabular}
\end{table}

\subsection{Alpha tagged source}\label{subsec:alpha}

\begin{figure}[thbp]
\includegraphics[width=\columnwidth]{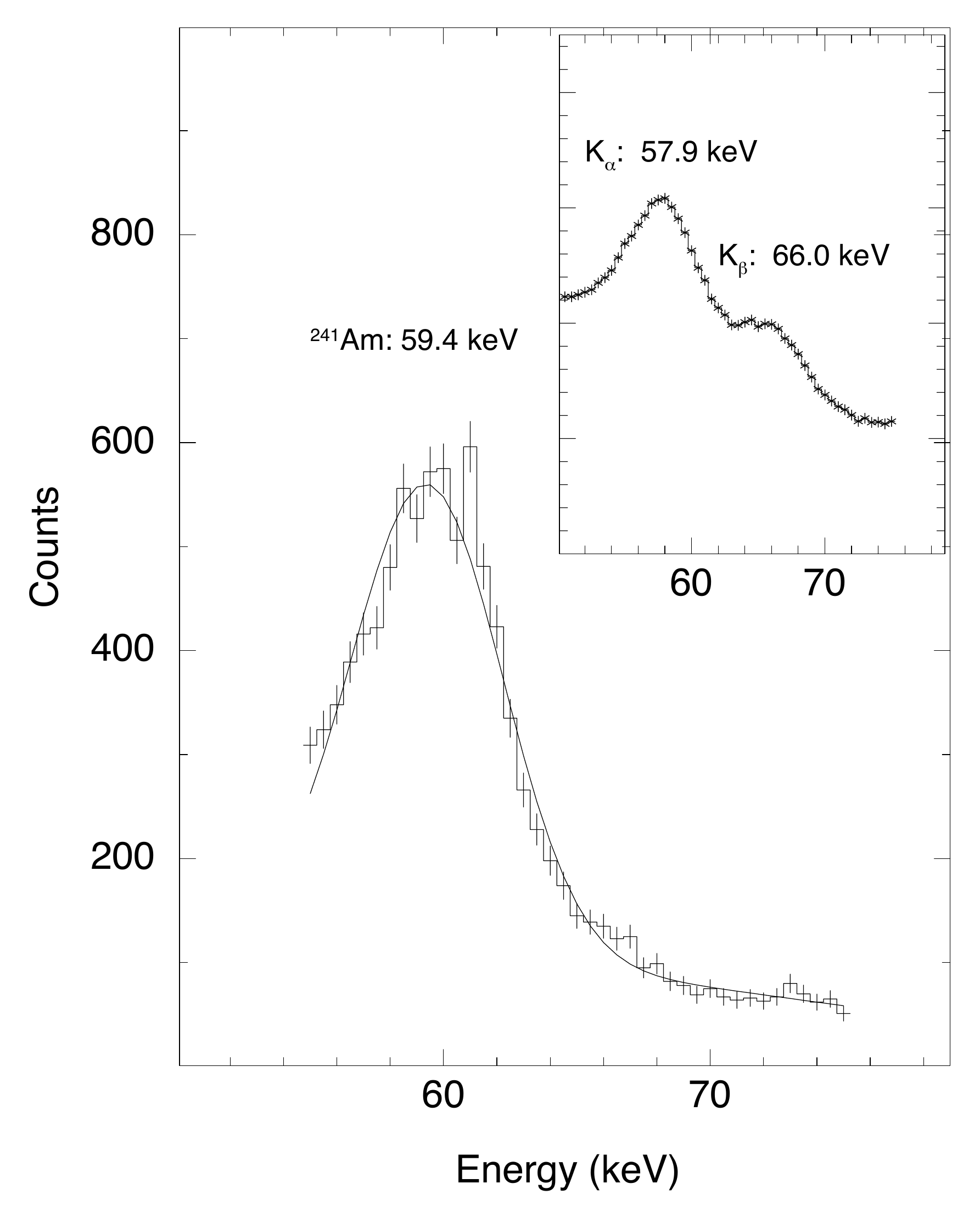}
\caption{In--orbit energy calibration spectra for Quadrant~A, extracted from data during observations of a faint source (\textless 150 mCrab).
The $\alpha$--tagged \element[241]{Am} source produces a line at 59.54~keV, which is recovered at $\sim59.4$~keV in this spectrum. 
 The inset shows background spectra with \element{Ta} fluorescence lines at 57.52 and 65.21~keV, recovered
at  $\sim$57.9~keV and $\sim$66.0~keV, respectively. }
\label{fig:alphaspec}
\end{figure}

For continuous calibration of gain and offset of the CZT Modules, a \element[241]{Am} source is kept 60~mm above and to the side of the detector plane (Figure~\ref{fig:cztilayout}). This source releases photons of 60~keV energy (Figure~\ref{fig:alphaspec}), with simultaneous emission of an alpha particle \citep{Rao2010}. The radioactive source is enclosed in a \element{CsI} (\element{Tl}) crystal, which detects the alpha particle and generates a photon signal. A photodiode coupled to this scintillator generates an electrical pulse, which is amplified in pre--amplifier and post--amplifiers. The post--amplifier signal is fed to an analog comparator along with lower-level discriminator level. 
Front end Electronics (\S\ref{subsec:electronics}) process the signal and set the alpha bit for any coincident \xray\ events. The coincidence test is implemented digitally as follows: the module response time is far faster than the alpha detectors, hence a delay $t_1$ (1~\us) is added to the arrival time of any \xray\ event\footnote{The $t_1$ delay is added only during onboard comparisons, and does not alter the actual timestamp of the \xray\ event.}. \xray s incident up to an interval $t_2$ (8~\us) after the alpha signal are considered as coincident with the alpha particle, and hence likely to be calibration photons. The electronics set the alpha bit to 1 for such \xray s. Finally, to avoid noise and multiple--counting, any further signals from the scintillator are ignored until a time $t_3$ (20~\us) elapses from the first scintillator signal.

In addition to the alpha--tagged source, \element{Ta} fluorescence lines also help in on--board energy calibration (Figure~\ref{fig:alphaspec}). These lines are fluorescent K$_\alpha$ and K$_\beta$ emissions arising from interaction between cosmic \xray\ photons and \element{Ta} in the CAM and collimators. The instrument team continuously monitors the Energy--to--PI conversion (gain and offset) of all modules using these lines.

\subsection{Veto detectors}\label{subsec:veto}
A veto detector placed 65.75~mm below the top surface of CZT modules serves as active shielding against charged particles giving \xray--like signatures in the detector, and background gamma-ray radiation (above 100~keV). The veto detector is a 20 mm thick \element{CsI} (\element{Tl}) scintillator, of size 167~mm $\times$ 167~mm, viewed by two one inch Photo-multiplier tubes (PMT) positioned at two sides of the crystal. These detectors are procured from Scionix Holland, the same suppliers who have provided the RT-2 phoswich detectors~\citep{dnr+2011}.

Photons normally incident on the CZT modules will get fully absorbed and will not penetrate to the veto detectors. Hence, any high energy photon passing through the veto detector is likely to be a background event. As in the case of alpha events, the coincidence is evaluated by the Front end Electronics Board using the timing parameters $t_1$ (1~\us\ delay to all CZT events) and $t_4$ (15~\us\ maximum delay between veto signal and CZT event). To suppress noise and ringing, any further signals from the veto detector are ignored for $t_5$ (40~\us). An additional complexity arises as the veto detector may also be triggered by charged particles. In this case, the width of the veto pulse is longer, and it continues to show spurious signals for a duration longer than $t_5$. This is fixed by having two more timing parameters: if the scintillator pulse width is longer than $t_6$ (20~\us), then the wait time is changed from $t_5$ to $t_7$ (496~\us) for that event. These timing parameters were determined from ground--based calibrations, and tweaked in--orbit to obtain the best background rejection.

The veto scintillator pulse is integrated and then digitised by an analog--to--digital convertor, to use as a proxy for energy of the incident photons. 
Any coincident \xray\ events in the CZT modules of that quadrant are tagged with the energy of the corresponding veto. Such events are transmitted to ground, but ignored by the default processing software. The FEB also creates a total spectrum of veto events for each quadrant every second. This spectrum is transferred to Processing Electronics for transmission to ground.

The veto detector is sensitive to photons with energies above 50~keV. The energy resolution of this detector is modest (about 20\%). Lab measurements show that the response of the detector as a function of energy is linear, and that the gain is uniform to $\sim10\%$ over the entire detector area. The amplifier gain has been adjusted so that the upper level corresponds to about 500--600~keV. The LLD, however, can be changed by command.

\subsection{Electronics configuration}\label{subsec:electronics}
Electronics of CZTI are designed at multiple levels. Each CZT module has its own on-board Application Specific Integrated Circuit (ASIC) with a FIFO buffer for storing events (\S\ref{subsec:modules}). In each quadrant, a Front end Electronics Board continuously polls each of the modules and reads out their signals. As mentioned in \S\ref{subsec:alpha} and \S\ref{subsec:veto}, the FEB checks for the coincidence of each \xray\ photon with signals from the alpha and veto detectors. The FEB prepares an event report for each detected event. The event reports prepared over a second are passed to the Processing electronics for transmission to ground.

The FEB contains high and low voltage supplies to provide the necessary power and biasing voltages to the detectors. All analog electronics are housed in the detector box itself, including ASIC control, current--to--voltage converters for all ASIC outputs, amplifiers etc. 

As discussed in \S\ref{subsec:modules}, \xray\ photon events are stored in the FIFO buffer of each module. Every 0.1~\us, the FEB polls two modules for events. If an event is present in the module's FIFO buffer, it takes 2.5~\us\ to read it. Note that this does not contribute to dead time, the module is capable of detecting other photons in this interval. The module provides pixel and PHA values for each event, to which the FEB adds module ID and timing information, and saves the time tagged event information in RAM. This timing information is accurate to 20~\us. If multiple events are present in the FIFO, only one is read --- the other events will be read in subsequent polling of the module. Thus, polling 16 modules, two at a time, takes anywhere between 0.8---20~\us\ depending on the total count rate. The FEB can store event reports for up to a total of 3072 events each second, any further events in that second are discarded.

The Processing Electronics (PE) is the master layer of electronics that forms the interface between the spacecraft and the CZTI Quadrants. The PE has built--in redundancy in the form of two identical fully functional sections, which can be switched from ground, so that it is not a single point of failure. Every second, the PE collects  time tagged data along with headers from all four FEBs in parallel. The on-board software of PE, which does the data handling operations, can be edited through commands from ground. PE software decides the mode of operation based on environmental conditions and ground commanded settings and packetises data in the respective format. The PE also has an interface with the GPS unit of the Satellite Positioning System (SPS), which provides a reference clock pulse with 200~ns accuracy every 16~s. This pulse is used to maintain absolute time reference for CZTI.

In orbit, it was observed that secondary photons from high energy photon/particle interactions with the spacecraft are a large source of background. Every such interaction creates a shower of photons over one or more detectors, creating even as high as 10--30 photon events within a single 20~\us\ time bin. Since these photons are of local origin, we have reprogrammed the PE to reject these photons on board. For every such ``bunch'' of photons, defined as 3 or more photons with $\le$20~\us\ separation between successive events, the PE re--encodes the data to give basic information about the nature of the bunch. In a typical observation, these reprocessed data are about five times smaller than the raw event data --- a significant saving in the downlink requirements.

Apart from packetising data, the PE also integrates housekeeping information from various sensors on board. These housekeeping data are packetised into Low Bitrate Telemetry (LBT) and is transmitted to Spacecraft on-board memory every second. The time tagged event data is packetised as High Bitrate Telemetry (HBT) and is transmitted to Spacecraft on-board memory every 21ms. The HBT and LBT data stored in Spacecraft on-board memory are transmitted to ground during the visibility of spacecraft to ground station. The PE receives time synchronisation pulses from the System Time Based Generator and provides the system time elapsed from the last pulse in the data frames packetised every second. The PE is responsible for receiving ground commands via the spacecraft, and transmitting them to respective instrument components. 

\section{Data}\label{sec:data}

\subsection{Data acquisition modes}
CZTI data can be acquired in various modes, based on a trade off between storage space (data volume) and data accuracy. The main modes used in regular observations are:
\begin{enumerate}
\item Normal mode (M0): This is the default operation mode of CZTI. Individual events are recorded with 10--bit energy information and 20~\us\ timing accuracy. The PE cleans any photon bunches (\S\ref{subsec:electronics}) and packetises all data for transmission to ground. Data frames are generated every second in this mode.
\item SAA mode (M9): When the spacecraft enters the South Atlantic Anomaly (SAA), CZTI high voltage is switched off for device safety. The passage into SAA can be detected automatically using counts from \asat's Charge Particle Monitor (CPM; Rao et. al., in prep), or pre--programmed by timing. In mode M9, the instrument only records house keeping, header and detector temperature information. In this mode, data frames are generated once in 100~seconds.
\item Secondary Spectral mode (SS): Secondary spectral mode data is acquired in parallel with the normal mode. This gives us integrated spectra of CZT, Veto and Alpha detectors, integrated over a duration of 100~s.
\end{enumerate}

In addition to this, there are several reduced data modes in the instrument which can be used to reduce the volume of data generated, at the same time, providing at least partial spectral or timing information. The transitions between modes can occur on ground command, or if certain pre-defined conditions are met onboard.
During the nine months of operation of the instrument, it was found that none of the reduced modes needed to be
activated and hence these modes are unlikely to be used during the operation of the instrument  in the future.

\subsection{Processing and products}
Control of overall \asat\ spacecraft as well as data down link from all scientific instruments is carried out from the ISRO Telemetry, Tracking and Command Network (ISTRAC) located at Bengaluru. The Payload data is then sent for processing at the payload operation centres (POC) for individual payloads and the higher level data is archived by Indian Space Science Data Centre (ISSDC) located at Bylalu. The payload operation centre (POC) for CZTI is located at the Inter University Centre for Astronomy \& Astrophysics (IUCAA) at Pune.

The CZTI POC processes all the ``Level~1'' data received from ISSDC to create astronomer--friendly ``Level~2'' products. The default pipeline filters data into different modes, identifies and suppresses noisy pixels, calculates livetime corrections (due to full data frames, see \S\ref{subsec:electronics}), creates images, as well as spectra and lightcurves for the primary on--axis source. The products are stored in \texttt{FITS} files compatible with the HEASARC \texttt{FTOOLS} package~\citep{blackburn95}. The data processing pipeline is compatible with the HEASoft package\footnote{The HEASoft software suite is maintained by the High Energy Archive Science Research Center (HEASARC), and can be downloaded freely from \url{http://heasarc.gsfc.nasa.gov/lheasoft/}.}. Various calibration files required for data processing are stored in the standard CALDB format. The CZTI processing pipeline, CALDB files, and sample data are all available in the ``Data and Analysis'' section of the the \asat\ Science Support Cell website, \url{http://astrosat-ssc.iucaa.in}.

Processed Level~2 data are validated by the POC and uploaded to ISSDC, where they can be downloaded by the proposers. As per \asat\ policy, data remain private for a period of one year from the date of observation, after which they are publicly accessible from the ISSDC website\footnote{ISSDC web site: \url{http://issdc.gov.in/astro.html}.}.

In order to ensure optimal use of CZTI's wide angle monitoring capabilities (\S\ref{subsec:offaxis}), the POC also undertakes the responsibility of scanning all data for astrophysical transients. Any transients detected in the data set are reported on the CZTI website\footnote{All transients observed by CZTI are listed at \url{http://astrosat.iucaa.in/czti/?q=grb}.} within one day. GCN circulars are also issued for the detected transients.

\subsection{In--orbit performance}\label{subsec:inorbit}
After its launch on 28 September 2015, the first six months of \asat\ were dedicated for performance verification observations, followed by a six--month long guaranteed time observation phase. \asat\ has observed several hard \xray\ sources like Crab, Cygnus~X--1, GRS~1915+105, Cygnus~X--3, etc. in this period. CZTI has also detected several GRBs and other hard X--ray transients. Some of the early results are reported in \citet{vrb+16}.

Observations of the Crab nebula were used for validating the mask and pixel alignment, timing accuracy, angular resolution, and overall response files. CZTI easily detected pulsations in the Crab pulsar, recovering a double--peaked pulse profile recorded by other hard \xray\ telescopes. Detailed analysis of simultaneous observations with GMRT (Radio) and {\em Integral} is under way.

\asat\ has undertaken several observations of Cygnus~X--1, giving good coverage of the transition from an ultra soft state to a hard state. Some of these observations had simultaneous coverage from {\em NuSTAR}, GMRT and the Mr. Abu Infra--Red Observatory, and joint analysis is under way.

As discussed in \S\ref{sec:intro}, CZTI can study hard \xray\ polarisation of sources. Hard \xray\ polarisation of Crab is clearly detected in CZTI data, and the team is carrying out refined analysis. Promising results have also been obtained for GRB160131A~\citep{vcm+16} and Cygnus~X--1.

\section{Conclusion}\label{sec:inorbit}
CZTI was the first science instrument to be made operational on \asat, on 6 October 2015. After initial tweaks to various settings, the instrument is now functioning perfectly, close to design specifications. 

The count rate of CZTI is background--dominated, and varies over \asat's orbit. The raw count rates in CZTI are around 1200--1300 counts/second in each quadrant. The on--board bunch clean algorithm was implemented on 16 February 2016, which reduced the Level~1 count rate to 400--450 counts/second in each quadrant. This algorithm intentionally uses conservative parameters to ensure that useful data is not discarded. Ground processing suppresses some noisy pixels, bringing the count rate down to around 100 counts/second in each quadrant. 

Several modules are showing low noise, and the median threshold (LLD) is set at about 22~keV. This provides a good overlap with the LAXPC energy range ($<80~keV$). Three of the 64 modules showed poor noise performance in--orbit, and their LLDs had to be raised above 60~keV, while three other modules operate with an LLD of $\sim50~keV$.

92\% of the 16,384 pixels are operational and can be used for imaging sources. About 77\% of the pixels have high spectroscopic quality. for these pixels, the resolution measured with the \element[241]{Al} source at 59.6~keV is $\lesssim 12\%$ (FWHM), with a median value of $10.8\%$. Less than 1\% pixels are noisy in a typical dataset, and are filtered out by the pipeline. These pixels often vary from orbit to orbit. If a pixel is consistently very noisy, commands are issued by the operators to disable it.

In a typical orbit, CZTI is inactive for about 20\% of the time due to SAA passage, and about 35\% is lost due to occultation of the source by the Earth. including other overheads, the net observing efficiency of CZTI is about 45\%.

Here we have provided an overview of the instrument and preliminary details of in--orbit performance. CZTI has observed several interesting hard \xray\ targets, results of these observations are being reported in the literature. Details of rigorous background modelling, spectral response, and imaging performance of the instrument will be presented elsewhere.

\section*{Acknowledements}

CZT--Imager is built by a consortium of Institutes across India. The Tata Institute of Fundamental Research, Mumbai, led the effort with instrument design and development. Vikram Sarabhai Space Centre, Thiruvananthapuram provided the electronic design, assembly and testing. ISRO Satellite Centre (ISAC), Bengaluru provided the mechanical design, quality consultation and project management. The Inter University Centre for Astronomy and Astrophysics (IUCAA), Pune did the Coded Mask design, instrument calibration, and Payload Operation Centre. Space Application Centre (SAC) at Ahmedabad provided the analysis software. Physical Research Laboratory (PRL) Ahmedabad, provided the polarisation detection algorithm and ground calibration. A vast number of industries participated in the fabrication and the University sector pitched in by participating in the test and evaluation of the payload.
The Indian Space Research Organisation funded, managed and facilitated the project.
This work utilised from various software including Python, IDL, FTOOLS, C, and C++.

\bibliographystyle{apj}
\bibliography{czti_v8.bib}

\end{document}